\documentclass[preprint2]{aastex}

\input epsf.sty

\slugcomment{To be published in the ApJL,
September 10, 2004}

\newcommand{\etal}{et~al.}
\newcommand{\eg}{e.g., }
\newcommand{\ie}{i.e., }
\newcommand{\Msun}{M_{\odot}}

\newcommand{\Fefs}{$^{56}$Fe}
\newcommand{\Cofs}{$^{56}$Co}
\newcommand{\Nifs}{$^{56}$Ni}
\newcommand{\Mej}{M_{\rm ej}}
\newcommand{\KE}{E_{\rm K}}
\def\gsim{\mathrel{\rlap{\lower 4pt \hbox{\hskip 1pt $\sim$}}\raise 1pt
\hbox {$>$}}}
\def\lsim{\mathrel{\rlap{\lower 4pt \hbox{\hskip 1pt $\sim$}}\raise 1pt
\hbox {$<$}}}

\begin{document}

\title{Supernova light curve models for the bump in the Optical Counterpart of X-Ray Flash 030723}

\author{
 N.~Tominaga\altaffilmark{1},
 J.~Deng\altaffilmark{1,2},
 P.A.~Mazzali\altaffilmark{1,2,3}
 K.~Maeda\altaffilmark{4},
 K.~Nomoto\altaffilmark{1,2},
 E.~Pian\altaffilmark{3},
 J.~Hjorth\altaffilmark{5}, and
 J.P.U.~Fynbo\altaffilmark{5,6}
 }

\altaffiltext{1}{Department of Astronomy, School of Science,
University of Tokyo, Bunkyo-ku, Tokyo 113-0033, Japan;
ntominaga@astron.s.u-tokyo.ac.jp, deng@astron.s.u-tokyo.ac.jp, nomoto@astron.s.u-tokyo.ac.jp}
\altaffiltext{2}{Research Center for the Early Universe, School of
Science, University of Tokyo, Bunkyo-ku, Tokyo 113-0033, Japan}
\altaffiltext{3}{INAF-Osservatorio Astronomico, Via Tiepolo, 11,
  34131 Trieste, Italy; mazzali@ts.astro.it, pian@bo.iasf.cnr.it}
\altaffiltext{4}{Department of Earth Science and Astronomy,
College of Arts and Science, University of Tokyo, Meguro-ku, Tokyo
153-8902, Japan; maeda@esa.c.u-tokyo.ac.jp}
\altaffiltext{5}{Niels Bohr Institute, Astronomical Observatory, University
of Copenhagen, Juliane Maries Vej 30, DK-2100
Copenhagen $\O$, Denmark; jens@astro.ku.dk}
\altaffiltext{6}{Department of Physics and Astronomy, University of
Aarhus, Ny Munkegade, DK-8000 ${\rm \AA}$rhus C, Denmark; jfynbo@phys.au.dk}

\begin{abstract}
XRF~030723 is the first X-ray flash (XRF) to show in its optical
light curve (LC) a bump that has been interpreted as the signature
of a supernova (SN). After subtracting the afterglow component
from the observed optical LC of the XRF counterpart, the properties
of the putative SN are constrained by means of synthetic LCs of
core-collapse SNe. For the redshift range
$z \sim 0.3$ -- 1, all possible models require a
rather small mass of synthesized \Nifs, \ie $M$(\Nifs) $\sim$ 0.01
-- 0.3 $\Msun$. The models used to describe the energetic SNe Ic
associated with gamma-ray bursts (SNe 1998bw and 2003dh) are too
massive for the observed LC. If the relation between ejected
\Nifs\ mass and total ejecta mass established from models of
various Type Ic SNe also holds for the putative SN in XRF~030723,
the ejecta mass is constrained to be $\sim$ 1 -- 3 $\Msun$ and the
kinetic energy $\lsim 1\times 10^{52}$ erg. This corresponds to a
progenitor with $15\Msun \lesssim M_{\rm MS}\lesssim 25\Msun$. The
SN therefore appears to have properties intermediate between a
normal SN~Ic like SN~1994I and a more energetic object like SN~2002ap.
\end{abstract}

\keywords{gamma rays: bursts -- supernovae: general --
nucleosynthesis -- X-rays: general}

\section{INTRODUCTION}

X-ray flashes (XRFs) are intense transient bursts of X-rays, similar to
gamma-ray bursts (GRBs) but characterized by a stronger X-ray than $\gamma$-ray
fluence \citep{hei01}. It has been speculated that XRFs are indeed linked to
GRBs. Various scenarios have been proposed: XRFs may be GRBs viewed
sufficiently off-axis (\citealt{yam03}), or more massive explosions than those
that make GRBs, leading to a smaller Lorenz factor for the jet
(\citealt{der00}). However, the real nature of the observed difference between
GRBs and XRFs is not yet known.

Both GRBs and XRFs are extragalactic sources. A number of GRBs
have now been localized, and they are all cosmological.
Interestingly, the three nearest long GRBs ever localized are all
associated with spectroscopically confirmed supernovae (SNe).
Although the case of GRB~980425/SN~1998bw \citep{gal98} was
controversial, that of GRB~030329/SN~2003dh is established beyond
doubt (e.g., \citealt{mat03}. GRB~031203 at $z = 0.1055$ also
appears to have hosted a SN possibly similar to SN~1998bw (e.g.,
\citealt{tho04}).

All these SNe are of Type Ic (SNe~Ic), and display very high-velocity
ejecta. SNe~Ic are thought to originate from the collapse of the
cores of massive stars that have lost both their H and He
envelopes, exploding as C+O stars (\eg \citealt{nom94}). SNe
1998bw and 2003dh have been successfully modelled as highly
energetic explosions (the spherical kinetic energy is $\sim 30 - 50$ times
that of a normal SN: \citealt{iwa98, maz03}), ejecting $\sim$ 10
$\Msun$ of material and synthesizing $\sim$ 0.5 $\Msun$ of \Nifs,
much more than in normal core-collapse SNe. The aspherical models
require lower energy, but still $2-10$ times higher than
a normal SN \citep{hof99, mae03}. These SNe~Ic and others,
not known to be associated with GRBs but showing the spectroscopic
signatures of a high explosion energy $(\gsim 10^{52}$ erg), have been
called hypernovae.  Both the kinetic energy
and the mass of \Nifs\ of SNe~Ic may positively
correlate with progenitor mass \citep{nom03}.

On the other hand, only a few XRFs have been accurately localized (to less than
a few arcmin) thus far, and only one of these is known to be at low redshift:
XRF~020903 at $z = 0.251$ \citep{sod03}.  In the case of XRF~030723, only upper
and lower redshift limits could be determined, because the host galaxy could
not be observed \citep{fyn04}. The lower limit, $z \gsim 0.3$, follows from the
non-detection of the host, while the upper limit, $z \lsim 2.3$, was derived
from the lack of Ly${\rm \alpha}$ absorption down to 4000  ${\rm \AA}$.

\citet{fyn04} obtained optical photometry and spectroscopy of XRF~030723. The
$R$-band light curve (LC) of the XRF counterpart showed a `bump', which may be
the signature of a SN component.  Other interpretations are possible (\eg a
two-jet model  \citealt{hua04}), but \citet{fyn04} claimed that they
could rule them out based on
the SED evolution. Interpreting the bump as a SN sets further limits on the
redshift ($z \sim 0.3 - 1.0$). \citet{fyn04} compared the bump with the LCs of
different SNe~Ic, concluding that the best match was given by the rest-frame
$B$-band LC of SN~1994I, at a redshift $z=0.6$. SN~1994I was a
normally-energetic SN~Ic, with low-mass ejecta \citep{nom94, iwa94}. However,
the featureless spectrum of the bump obtained by \citet{fyn04} makes them
favour a broad-lined, highly energetic SN, like SN~1998bw or SN~2002ap.

In this paper, we model the observed LC using different SN~Ic models,
both normally- and hyper-energetic, and explore various redshifts to
determine the possible range of the parameters of the SN associated with
XRF~030723.

\section{Light Curve Models}

Assuming that the SN in XRF~030723 is of Type Ic, by analogy with
the cases of GRB-associated SNe, we computed synthetic $UVOIR$
bolometric LCs of exploded C+O star models. We used an LTE
radiation hydrodynamical code and a gray $\gamma$-ray transfer
code \citep{iwa00}. The electron-scattering opacity was calculated
from the solution of the Saha equation, while the line opacity was
fixed at 0.03 ${\rm cm^2~g^{-1}}$, the value that was used to
model SN~1998bw \citep{nak01}. The total opacity is $\sim 0.03$ --
0.1 ${\rm cm^2~g^{-1}}$ (compared to the range 0.05 -- 0.15
${\rm cm^2~g^{-1}}$ for SN~1994I: \citealt{iwa94}).
The accurate line opacity is unknown, if it is bigger,
the timescale of LC peak becomes longer.

We started from four different C+O star explosion models: CO21, a
normal SN Ic model developed for SN~1994I \citep{iwa94}; CO100, an
energetic model developed for SN~1997ef \citep{maz00}; CO100/4, a
scaled-down version of CO100 developed for SN~2002ap
\citep{maz02}; and CO138, a model for the very energetic SN~1998bw
\citep{nak01}. The model parameters, i.e. ejected mass $M_{\rm
ej}$, kinetic energy $\KE$, and progenitor main-sequence mass
$M_{\rm MS}$ are summarized in Table \ref{omodel}.

Given the uncertain redshift, we considered three values:
$z=0.3$, 0.6, and 0.8. The observed $R$-band roughly corresponds
to the rest-frame $V$-band if $z=0.3$, the $B$-band if $z=0.6$,
and the $U$-band if $z=0.8$. Because of
time-dilation, the observed LC corresponds to narrower rest-frame
LCs for increasing redshifts. On the other hand, in SNe~Ic the
LC peak is narrower for bluer bands. Coincidentally, these two
effects roughly cancel out.

We scaled the masses of the above C+O models, conserving their
explosion energies, and calculated synthetic LCs to find which
scalings yield the best-fit models. Since the code computes
bolometric LCs, the monochromatic LCs were estimated using
bolometric corrections (BCs). As spectral and color information
about the possible SN is very limited, we used the BCs of known
SNe Ic as templates. These were calibrated in time with respect to
the LC peak to match the apparently fast evolution of the SN. We
used the BC template appropriate for the parameters of each model.
For low-mass, low-energy models, we used the BCs of SN~1994I
(assuming $E(B-V)=0.45$; \citealt{ric96}; B. Schmidt \& R. Kirshner 1994,
private communication). In other cases, the model parameters
indicated energetic explosions, and then we adopted the BCs of a suitable
hypernova and recalculated the model LC. In order to match the observed
flux, we modified the mass of \Nifs\ synthesized for different
redshifts and models. \Nifs\ powers the SN LC through the decay
chain \Nifs\ $\rightarrow$ \Cofs\ $\rightarrow$ \Fefs.

Model parameters affect the synthetic LC. The
timescale of the LC peak, $\tau_{\rm peak}$, is determined by the
combination of photon diffusion and ejecta expansion and depends
on ${\KE}$ and $M_{\rm ej}$ as $\tau_{\rm peak}
\approx A {\kappa_{\rm opt}}^{1/2} {M_{\rm
ej}}^{3/4}{{\KE}}^{-1/4}$ \citep{arn82}, where $A$ represents the
effect of the density structure and \Nifs\ distribution and
$\kappa_{\rm opt}$ is the model-dependent average opacity.

\begin{figure*}
\centering
\epsscale{2.5}
\plotone{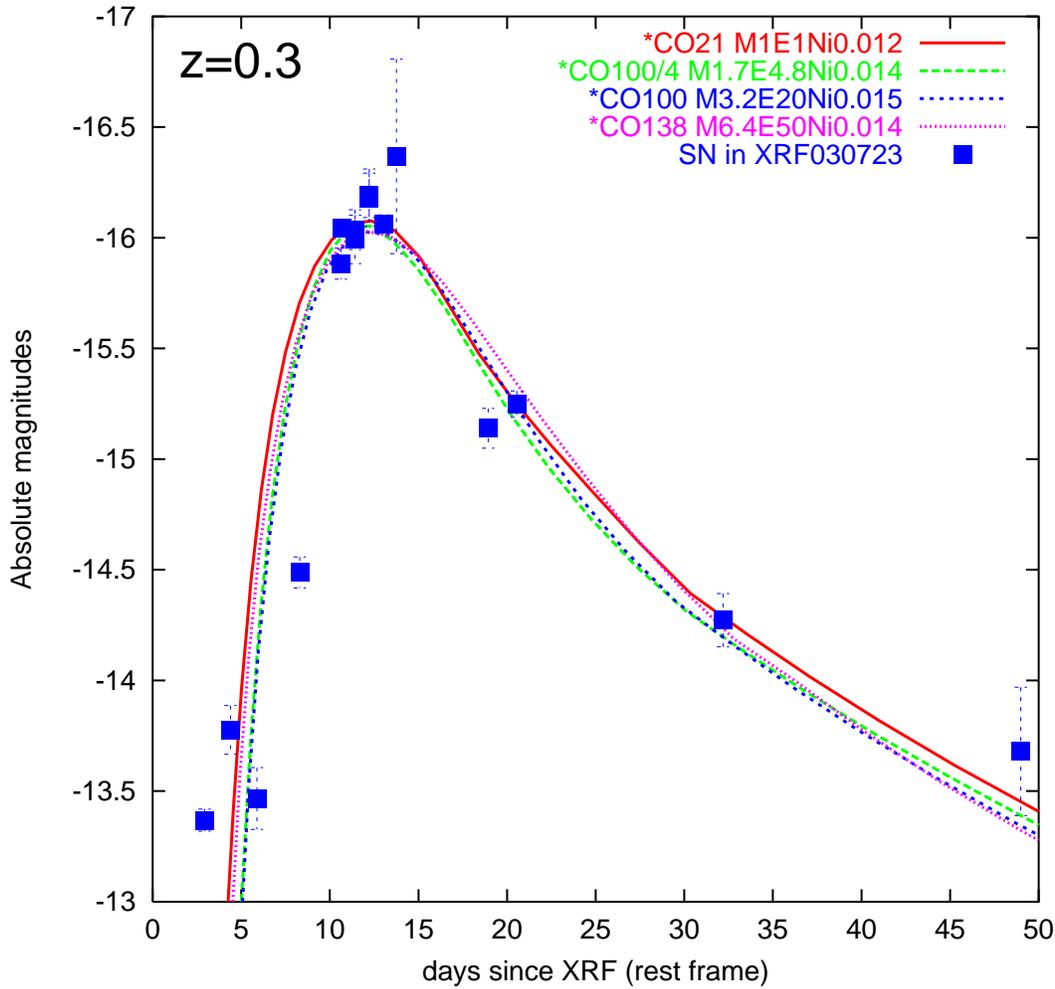} \figcaption[f1.eps] {Comparison
between the best-fit model LCs ({\em solid line}: $^*$CO21; {\em
long-dashed line}: $^*$CO100/4; {\em short-dashed line}: $^*$CO100; {\em
dotted line}: $^*$CO138) and the observed LC for z=0.3 ({\em filled
squares}; \citealt{fyn04}). \label{z03}}
\end{figure*}

\begin{figure*}
\centering
\epsscale{2.5}
\plotone{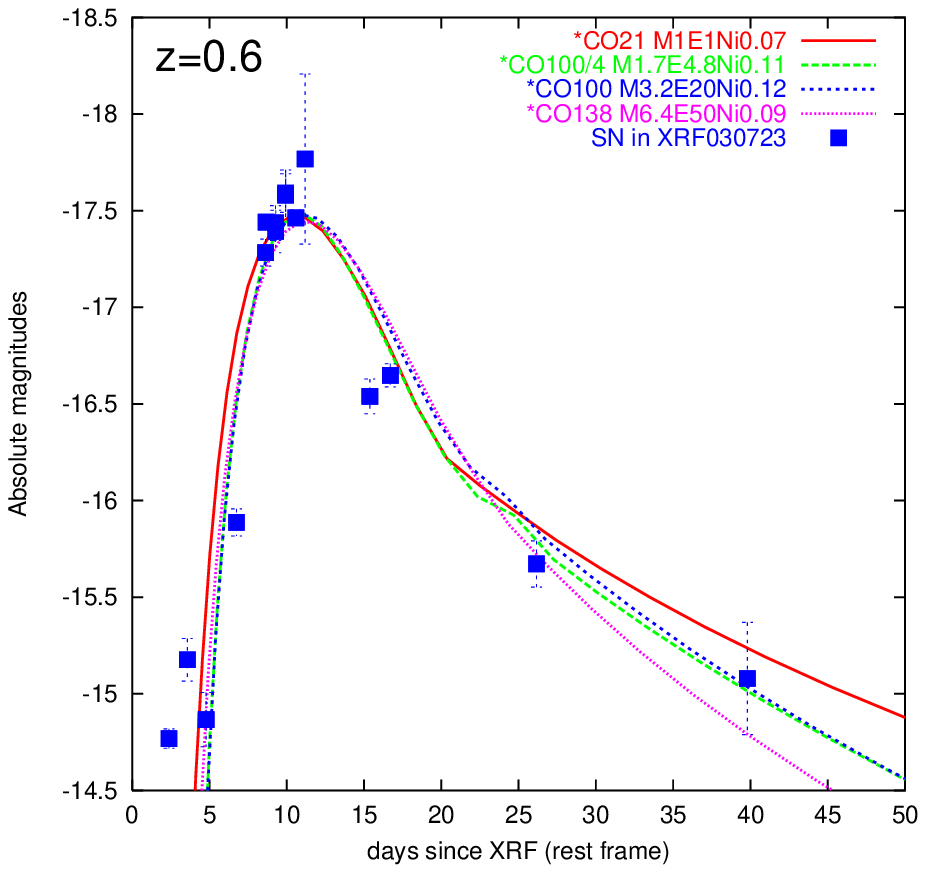} \figcaption[f2.eps] {Same
as Fig. 1, but for z=0.6. \label{z06}}
\end{figure*}

\begin{figure*}
\centering
\epsscale{2.5}
\plotone{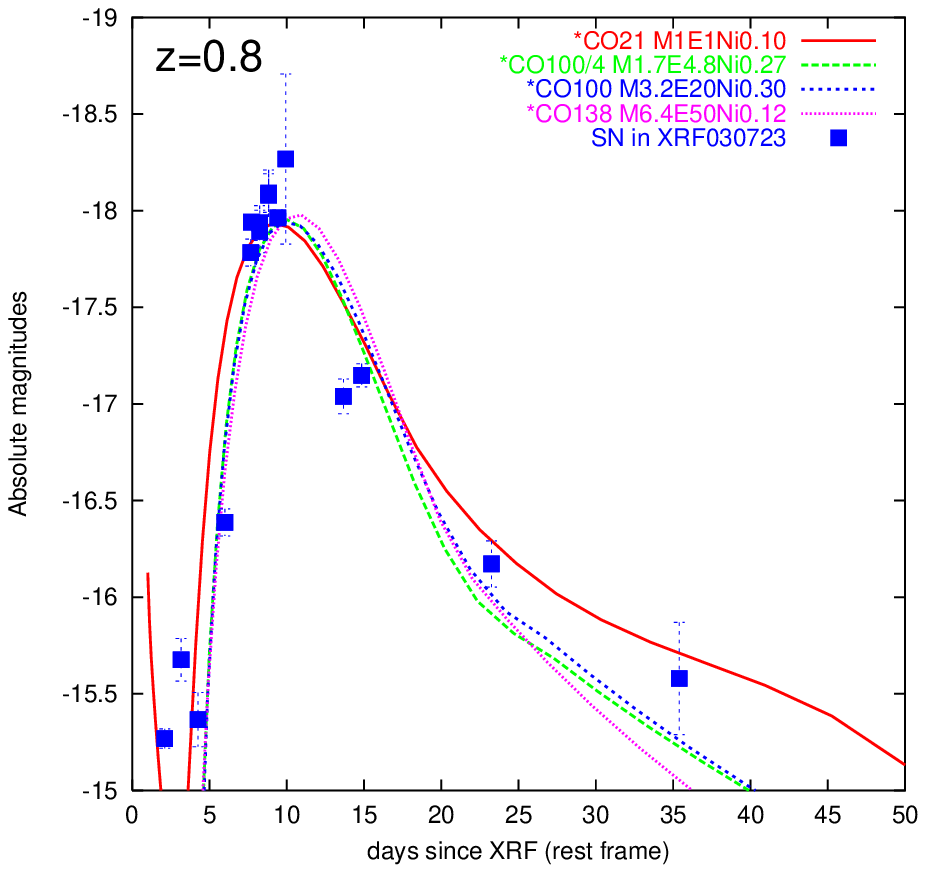} \figcaption[f3.eps] {Same
as Fig. 1, but for z=0.8. \label{z08}}
\end{figure*}

The parameters of our best-fit models are listed in Table~\ref{model}, as are
the template BCs used, and the computed synthetic LCs are compared to the
observed LC in Figures \ref{z03}, \ref{z06}, and \ref{z08} for the three
redshifts.  The observed LC was derived from the \citet{fyn04} R-band
photometry by subtracting the Beuermann extrapolation of the power-law
component of the afterglow. As expected, for every value of ${\KE}$ (\ie for
every model) there is a value of the ejected mass for which the synthetic LC
reproduces the observations. Had we been able to model the spectra, we could
have uniquely constrained the value of the kinetic energy. However, the spectral
information is limited. \citet{fyn04} discussed that the SED of the bump is
consistent with that of a SN similar to SN~1998bw, and that the only, low
signal-to-noise spectrum is better consistent with a broad-lined SN like
SN~1998bw or SN~2002ap than a narrow-lined one like SN~1994I. We discuss these points later.

Because the LC peak of the SN in XRF~030723 is intrinsically
narrow, the only models that give satisfactory results without
major modifications to their mass values are models CO21, which is
scaled up somewhat, and CO100/4, which has to be scaled down (the
top panel of Figure~\ref{em}). All other models were originally
developed for the high-mass HNe and must be scaled down in mass
significantly to yield a narrow LC peak.

The delayed LC-rise of the SN in XRF~030723, compared with
the hyper-energetic SNe 1998bw and 2002ap (\eg
\citealt{gal98,yos03}), cannot be reproduced by models with
uniform \Nifs\ mixing. Instead, we had to assume that \Nifs\ is
distributed only in the inner 10\% of the ejected mass in our
modelling. This increases the photon diffusion time, leading to a
delayed onset of the LC. The total \Nifs\ mass is $\sim$ 0.012 --
0.015 $\Msun$ if $z = 0.3$, $\sim$ 0.07 -- 0.12 $\Msun$ if $z =
0.6$, and $\sim$ 0.1 -- 0.3 $\Msun$ if $z = 0.8$. These estimates
would change if we could take into account the unknown
$K$-corrections. For example, the $K$-correction of SNe Ia between
the observed $R$- and rest-frame $B$-bands, $K_{BR}$, at $z = 0.6$
is $\sim -0.6$ \citep{kim96}. If a similar value was applied here
(SNe Ic show some overall spectral similarity to SNe Ia), the
\Nifs\ mass for $z = 0.6$ would decrease to $\sim$ 0.04 $\Msun$.
In the case $z = 0.3$ the relevant $K$-correction for SNe Ia,
$K_{VR}$, varies between $-0.2$ and $-0.5$ near the peak, which
suggests that the \Nifs\ mass could decrease to $<$ 0.01 $\Msun$.
The $K$-correction for $z = 0.8$, $K_{UR}$, is possibly positive
for SNe Ic because of the very small flux in the $U$-band. This
could make the \Nifs\ mass larger than 0.25 $\Msun$.

\section{Discussion}

All the models discussed above can reproduce the observed LC.
Given the limited spectral information, it is unfortunately not
easy to select among them. However, based
on their properties, we can at least attempt to narrow down the
range of possibilities.

\begin{figure*}
\centering
\epsscale{2.5}
\plotone{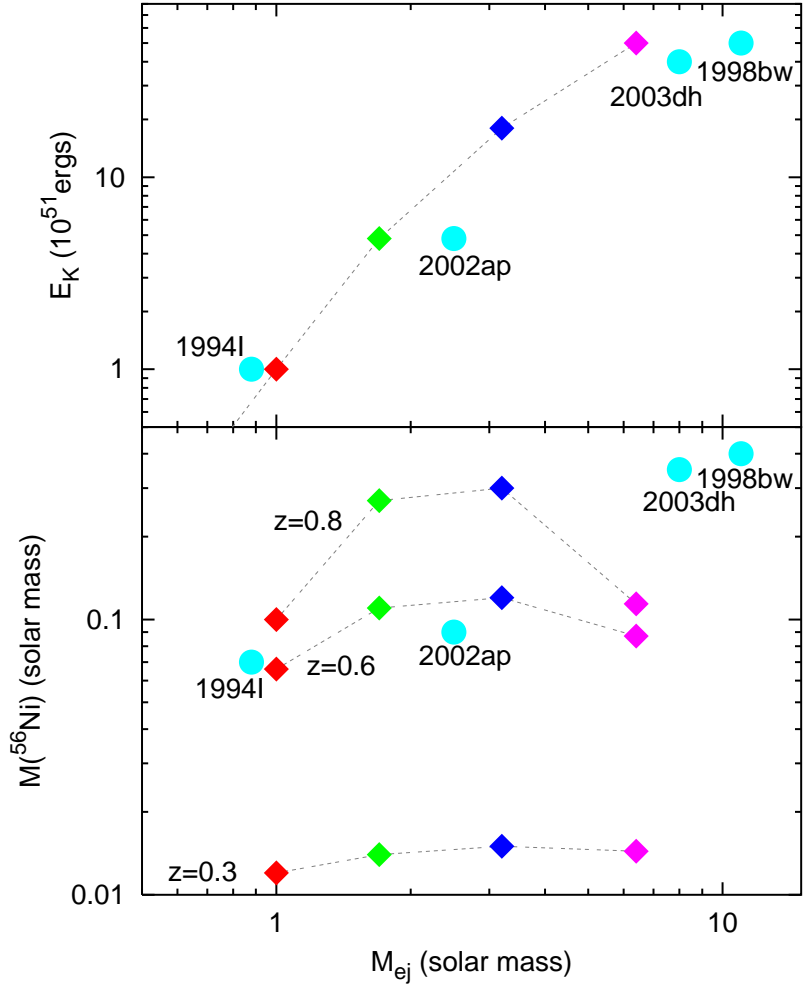} \figcaption[f4.eps] {{\em Top panel}:
$M_{\rm ej}$ vs. ${\KE}$ for the best-fit models ({\em squares}) and
for four well-studied SNe~Ic ({\em filled circles}). {\em
Bottom panel}: $M_{\rm ej}$ vs. $ M{\rm(^{56}Ni)}$ for the
best-fit models ({\em squares}) at $z=0.8$ ({\em top}), 0.6 ({\em
middle}), and 0.3 ({\em bottom}), respectively, and for the
SNe~Ic ({\em filled circles}). \label{em}}
\end{figure*}

In Figure~\ref{em}, the basic parameters of the best-fit models
are compared with the properties of four well-studied SNe~Ic of
various energies. The {\em top panel} is a plot of $\Mej$ vs. $\KE$ for the
various models and SNe. The {\em bottom panel} shows how $M({\rm
^{56}Ni})$ varies depending on the assumed redshift, all other
model properties being the same. Redshift affects both the
estimated SN luminosity and the BCs adopted in the LC modelling.
Based on this plot, we can select the models that are likely to be
applicable to the SN in XRF~030723, assuming that it is a SN Ic.
The models for the observed Type Ic SNe are also plotted.

First we discuss the low-redshift case ($z=0.3$). In this case,
the mass of \Nifs\ is very small ($\lesssim 0.015 \Msun$) for all
models. This value is much smaller than even in a normal SNe~Ic like
SN~1994I, suggesting that $z=0.3$ is a significant underestimate
of the real redshift if the observed bump is in fact a SN.
However, if XRF~030723 really occurred at
$z=0.3$, then strong fall-back and/or a low-energy SN must be
invoked to explain such a small \Nifs\ mass. If fall-back is
strong, much of the synthesized \Nifs\ may be captured by the
compact remnant, like in some Type II SNe which eject very little
\Nifs\ (\eg SN~1997D, \citealt{tur98,zam03}). The other
possibility is that the explosion energy of the SN is too low to
synthesize much \Nifs. The ${\KE}$ -- $M_{\rm ej}$ relation of our
best-fit models suggests that SNe with $(M_{\rm ej},~{\KE})$
as low as, \eg $(0.8,~0.5)$ would have the correct LC and may be
expected to synthesize little \Nifs\, as is required for $z=0.3$.
Such SNe must come from low-mass C+O stars, originating from
main-sequence stars of $< 15 \Msun$.

Next we consider the high-redshift case ($z=0.8$). This sets an
upper limit for the ejected \Nifs\ mass at $M({\rm ^{56}Ni}) \sim$
0.3 $\Msun$. Combined with the tendency of actual SNe Ic to
produce more \Nifs\ for increasing $\Mej$ and $\KE$, this in turn
sets an upper limit to $\Mej$ and ${\KE}$. The ejecta mass of the
SN in XRF~030723 is probably less than  $\sim 6 \Msun$ (\ie
$M_{\rm MS} < 30$ $\Msun$), and the kinetic energy less than $\sim
3 \times 10^{52}$ erg. These values are smaller than in the
GRB-associated SNe. It is very unlikely that the SN can be as
massive as SNe 1998bw \citep{iwa98} or 2003dh \citep{maz03} for
two reasons: first, the kinetic energy would be extremely high
($\sim 1 \times 10^{53}$ erg), and secondly, it would be strange
for such a hyper-energetic explosion only to synthesize $\sim$ 0.2
$\Msun$ of \Nifs.

If the redshift $z$ is $\sim$ 0.6, the required \Nifs\ mass is
$\sim 0.07 - 0.12 \Msun$. Both the SN~1994I-like model $^*$CO21
($\Mej\sim 1 \Msun$, ${\KE}\sim 1 \times 10^{51}$ erg,
$M$(\Nifs) $\sim 0.07 \Msun$) and the SN~2002ap-like model
$^*$CO100/4 ($\Mej\sim 1.7 \Msun$, ${\KE}\sim 3 \times 10^{51}$ erg,
$M$(\Nifs) $\sim 0.1 \Msun$) are consistent with the observed
relations between $\Mej$, ${\KE}$, and $M$(\Nifs). If the redshift
is slightly higher (\eg $z \sim 0.7$), a larger $M$(\Nifs) is
required, and a more massive and energetic model ($\Mej\sim 2 - 3
\Msun$; ${\KE} \sim 4$ -- $10 \times 10^{51}$ erg) is probably
necessary. This makes the tentative SN more like SN~2002ap.  Summarizing, our preferred estimates
are for a redshift $z \sim 0.5 - 0.7$, consistent with the
range favored by \citet{fyn04}, and SN properties between a normal
SN and a low-energy, low-mass hypernova.

These results are consistent with the colors of the putative SN
on August 6 and 19, \ie 14 and 21 days after the XRF \citep{fyn04}.
We made a rough comparison between the XRF and the color
evolution of other Type Ic SNe for the three redshifts,
neglecting the unavailable K-corrections. The observed $V-R \sim
0.1- 0.25$ and $i-R \sim -0.4 - -0.2$ on August 6 cannot be used
to distinguish between the three redshifts and between SNe 1998bw,
2002ap, and 1994I.  However, the $UB$ non-detection on August 16,
($B-R>2$ and $U-R>2$) seems to match SN 2002ap better
because the $U/UV$ flux is more depressed in SN 2002ap
than in SNe 1994I or 1998bw. The detected $K=21.2$ on Aug 14, however,
suggests $R-K\sim 3$, which is too big to be consistent with the
photometry of SNe 1998bw, 2002ap, and 1994I (but see \citealt{fyn04}).
Finally, the observed $i-R \sim -1.1$ on
August 16 favors $z \sim 0.6$ and a comparison with SNe 1994I and 2002ap.
Support for an SN~2002ap-like object also comes from the apparently featureless
spectrum of the bump \citep{fyn04}. SN~2002ap has a broad-lined spectrum, much
more similar to SN~1998bw than to SN~1994I.

This is the first case of an XRF probably associated with a SN \citep{wat04}.
%(For the X-ray rich GRB~020410, \citet{lev04} reported the
%possible association with a SN of low \Nifs\ mass.)
Interestingly, the properties of the SN appear to lie at the low
end of the distribution of HN properties, or perhaps even to be
similar to those of normal SNe~Ic (Figure~\ref{em}). This may be
the consequence of an unfavorable orientation which prevented us
from seeing the GRB and the most energetic part of the ejecta,
but it also may indicate a real difference between XRFs and GRBs
on the one hand and the accompanying SNe on the other. If the
progenitor of the SN/XRF is really intermediate between those of SN~2002ap and
1994I, \ie $15\Msun \lesssim M_{\rm MS}\lesssim 25\Msun$, a black
hole may not be formed (\eg if $M_{\rm MS}\lesssim 20\Msun$ as in
\citealt{fry01}), and the central engine may be a neutron star
like in the ``magnetar'' model \citep{nak98}, instead of the
black hole of the ``collapsar'' model for long GRBs
\citep{mac99}. Accurate spectral information, which would have
allowed us to constrain the properties of the SN much more
tightly, is unfortunately not available. There is however one case
where a normal, or possibly SN~2002ap-like SN~Ic was claimed to be
associated with a GRB (GRB~021211/SN~2002lt, \citealt{del03}).
Clearly, more data on the SN -- GRB/XRF connection are
necessary before we can understand the full extent of the relation
between these phenomena.

\clearpage

\begin{deluxetable}{clcccc}
 \tabletypesize{\scriptsize}
 \tablecaption{Original SN/HN Models\label{omodel}}
 \tablewidth{0pt}
 \tablehead{
   \colhead{Model}
 & \colhead{Supernova}
 & \colhead{$M_{\rm MS}/\Msun$}
 & \colhead{$M_{\rm ej}/\Msun$}
 & \colhead{${\KE}$}\\
   \colhead{}
 & \colhead{}
 & \colhead{}
 & \colhead{}
 & \colhead{($10^{51}$ ergs)}
 }
\startdata
CO21       & SN~1994I  & $\sim$ 15 & 0.88 & 1\\
CO100/4    & SN~2002ap & 20 -- 25 & 2.4  & 4.8\\
CO100      & SN~1997ef & 35 -- 40 & 9.6  & 20\\
CO138      & SN~1998bw & $\sim$ 40  & 11   & 50\\
\enddata
\end{deluxetable}

\begin{deluxetable}{ccccccc}
 \tabletypesize{\scriptsize}
 \tablecaption{Best-fit Models\label{model}}
 \tablewidth{0pt}
 \tablehead{
   \colhead{Model \tablenotemark{a}}
 & \colhead{$M_{\rm ej}/\Msun$}
 & \colhead{${\KE}$}
 & \colhead{}
 & \colhead{$M$(\Nifs)/$\Msun$}
 & \colhead{}
 & \colhead{Bolometric Correction}\\
   \colhead{} &
   \colhead{} &
   \colhead{($10^{51}$ ergs)} &
   \colhead{$z=0.3$} &
   \colhead{$z=0.6$} &
   \colhead{$z=0.8$} &
   \colhead{Template \tablenotemark{b}}
 }
\startdata
$^*$CO21 & 1 & 1 & 0.012 & 0.07 & 0.10 & SN~1994I\\
$^{*}$CO100/4 & 1.7 & 4.8 & 0.014 & 0.11 & 0.27 & SN~2002ap \\
$^{*}$CO100 & 3.2 & 20 & 0.015 & 0.12 & 0.30 & SN~2002ap \\
$^{*}$CO138 & 6.4 & 50 & 0.014 & 0.09 & 0.12 & SN~1998bw \\
\enddata
\tablenotetext{a}{The subscript $^*$ is used to discriminate each
modified model from the original model.}

\tablenotetext{b}{The BCs for SNe~1998bw and 2002ap are scaled up
in time by a factor of 1.5 and 1.2, respectively, with respect to
the peak epoch.}
\end{deluxetable}

\clearpage

\clearpage 

\clearpage 

\clearpage

\begin{thebibliography}{}

\bibitem[Arnett(1982)]{arn82} Arnett, W. D. 1982, \apj, 253, 785

%\bibitem[Cobb et al.(2004)]{cob04} Cobb, B., Bailyn, C., van
%    Dokkum, P., Buxton, M., \& Bloom, J. 2004, \apj, 608, L93

\bibitem[Della Valle \etal(2003)]{del03} Della Valle, M., \etal\
2003, \aap, 406, L33

\bibitem[Dermer, Chiang, \& Mitman(2000)]{der00} Dermer, C.D., Chiang,
              J., \& Mitman, K.E. 2000, \apj, 537, 785

\bibitem[Fryer \& Kalogera(2001)]{fry01} Fryer, C. L., \&
Kalogera, V. 2001, \apj, 554, 548

\bibitem[Fynbo \etal(2004)]{fyn04} Fynbo, J. P. U., \etal\ 2004,
\apj, in press (astro-ph/0402240)

\bibitem[Galama \etal(1998)]{gal98} Galama, T. J., \etal\ 1998,
\nat, 395, 670

%\bibitem[Gal-Yam \etal(2004)]{gal04} Gal-Yam, A., \etal\ 2004,
%\apj, 609, L59

\bibitem[Heise \etal(2001)]{hei01} Heise, J., in't Zand, J.,
Kippen, R., \& Woods, P. 2001, in GRBs in the
Afterglow Era, 16

%\bibitem[Hjorth \etal(2003)]{hjo03} Hjorth, J., \etal\ 2003,
%\nat, 423, 847

\bibitem[H\"oflich \etal(1999)]{hof99} H\"oflich, P., \etal\ 1999,
\apj, 521, 179

\bibitem[Huang \etal(2004)]{hua04} Huang, Y.F., \etal\ 2004,
\apj, 605, 300

\bibitem[Iwamoto \etal(1994)]{iwa94} Iwamoto, K., Nomoto, K.,
H\"{o}flich, P., Yamaoka, H., Kumagai, S., \& Shigeyama, T. 1994,
\apj, 437, L115

\bibitem[Iwamoto \etal(1998)]{iwa98} Iwamoto, K., \etal\ 1998,
\nat, 395, 672

\bibitem[Iwamoto \etal (2000)]{iwa00} Iwamoto, K., \etal\ 2000,
\apj, 534, 660

%\bibitem[Kawabata \etal(2003)]{kaw03} Kawabata, K. S., \etal\
%2003, \apj, 593, L19

\bibitem[Kim, Goobar, \& Perlmutter(1996)]{kim96} Kim, A., Goobar, A., \&
Perlmutter, S. 1996, \pasp, 108, 190

%\bibitem[Levan \etal(2004)]{lev04} Levan, A., \etal\ 2004, \apj,
%submitted (astro-ph/0403450)

\bibitem[MacFadyen \& Woosley(1999)]{mac99} MacFadyen, A. I., \&
Woosley, S. E. 1999, \apj, 524, 262

%\bibitem[Malesani et al.(2004)]{mal04} Malesani, D., et al. 2004,
%    \apj, 609, L5

\bibitem[Matheson \etal (2003)]{mat03} Matheson, T., \etal\
2003, \apj, 599, 394

\bibitem[Maeda \etal (2003)]{mae03} Maeda, K., Mazzali, P.A., Deng, J.,
              Nomoto, K., Yoshii, Y., Tomita, H., \& Kobayashi, Y.
              2003, \apj, 593, 931

\bibitem[Mazzali, Iwamoto, \& Nomoto(2000)]{maz00} Mazzali, P.A.,
Iwamoto, K., \& Nomoto, K. 2000, \apj, 545, 407

\bibitem[Mazzali \etal(2002)]{maz02} Mazzali, P. A., \etal\ 2002, \apj,
572, L61

\bibitem[Mazzali \etal(2003)]{maz03} Mazzali, P. A., \etal\ 2003,
\apj, 599, L95

\bibitem[Nakamura \etal (1998)]{nak98} Nakamura, T. 1998, Prog. Theor. Phys., 100,
921

\bibitem[Nakamura \etal (2001)]{nak01} Nakamura, T., Mazzali, P. A.,
Nomoto, K., Iwamoto, K. 2001, \apj, 550, 991

\bibitem[Nomoto \etal(1994)]{nom94} Nomoto, K., Yamaoka, H.,
Pols, O. R., van den Heuvel, E. P. J., Iwamoto, K., Kumagai, S.,
\& Shigeyama, T. 1994, \nat, 371, 227

\bibitem[Nomoto \etal(2003)]{nom03} Nomoto, K., Maeda, K,
Mazzali, P. A., Umeda, H., Deng, J., \& Iwamoto, K. 2004, in
Stellar Collapse, ed. C. L. Fryer (Dordrecht: Kluwer), in press
(astro-ph/0308136)

%\bibitem[Prochaska \etal(2004)]{pro04} Prochaska, J. X., \etal\
%2004, \apj, in press (astro-ph/0402085)

\bibitem[Richmond \etal(1996)]{ric96} Richmond, M. W., \etal\
1996, \aj, 111, 327

%\bibitem[Stanek \etal(2003)]{sta03} Stanek, K. Z., \etal\ 2003,
%\apj, 591, L17

\bibitem[Soderberg \etal(2003)]{sod03} Soderberg, A. M., \etal\
2003, \apj, submitted (astro-ph/0311050)

\bibitem[Thomsen \etal(2004)]{tho04} Thomsen, B., \etal\ 2004,
\aap, 419, L21

\bibitem[Turatto \etal(1998)]{tur98} Turatto, M., \etal\ 1998,
\apj, 498, L129

\bibitem[Watson \etal(2004)]{wat04} Watson, D., \etal\ 2004,
\apj, 605, L101

\bibitem[Yamazaki \etal(2003)]{yam03} Yamazaki, R., Ioka, K., \&
Nakamura, T. 2003, \apj, 593, 941

\bibitem[Yoshii \etal(2003)]{yos03} Yoshii, Y., \etal\ 2003,
\apj, 592, 467

\bibitem[Zampieri \etal(2003)]{zam03} Zampieri, L., \etal\ 2003,
\mnras, 338, 711

\end{thebibliography}
\end{document}